\documentclass[11pt,a4paper]{article}
\usepackage[margin=1in]{geometry}
\usepackage{setspace}
\usepackage{graphicx} 
\usepackage{amsmath, amssymb, amsthm}
\usepackage{appendix}

\title{A Quantum Interface Between Neutral-Atoms and Trapped-Ions
Quantum Registers}
\author{
Ayelet Hasson$^{1}$,
Gal Dekel$^{1}$,
Ehud Shahar$^{1}$,
Nitzan Akerman$^{1}$,
Roee Ozeri$^{1}$\\[1ex]
\small $^{1}$Department of Physics of Complex Systems, Weizmann Institute of Science, Rehovot 7610001, Israel
}
\date{July 2026}

\begin{document}

\maketitle

\begin{abstract}
Hybrid quantum systems combining neutral atoms and trapped ions offer the prospect of integrating the scalability of atom arrays with the high-fidelity control available in trapped-ion platforms. Here we propose and analyze a quantum interface between individually trapped neutral atoms and a trapped-ion crystal. In our scheme, a neutral $^{88}$Sr atom trapped in optical tweezers interacts with a small $^{88}$Sr$^+$ ion crystal, and by exciting the atom to a Rydberg state, the atom-ion polarization interaction is strongly enhanced, resulting in a state-dependent modification of the ions' collective motional modes. We show that this shift enables conditional control of a M{\o}lmer--S{\o}rensen gate, allowing the neutral atom to act as a control qubit for an entangling operation between two ions. We investigate the feasibility of the scheme by analyzing Rydberg trapping in the combined optical tweezers and Paul trap potentials, identifying negative-polarizability Rydberg states as particularly favorable for stable confinement. We further evaluate the relevant trapping conditions, Rydberg lifetimes, and coherence requirements, and show that the proposed interface is compatible with realistic experimental parameters. These results establish a practical route toward deterministic atom--ion hybrid quantum gates and quantum interfaces.
\end{abstract}

\section{Introduction}
\label{sec-Introduction}\

Neutral atoms and trapped ions are among the leading platforms for quantum computing~\cite{1-Bluvstein_2022,2-Liu_2025}, quantum simulation~\cite{3-Georgescu_2014}, and optical clocks~\cite{4-RevModPhys.87.637}. Neutral-atom systems offer excellent scalability and flexible geometries enabled by optical tweezers arrays and Rydberg interactions~\cite{1-Bluvstein_2022,42-PhysRevLett.85.2208,43-RevModPhys.82.2313,44-doi:10.1126/science.aah3778, 45-Bernien_2017,46-Browaeys_2020,47-Bluvstein_2023}, while trapped ions provide long coherence times and exceptionally high-fidelity quantum gates mediated by collective motional modes~\cite{7-hughes2025trappediontwoqubitgates9999, 48-Leibfried2003ExperimentalDO, 49-PhysRevLett.117.060504, 50-Wang_2017}. At the same time, neutral-atom platforms are typically limited by lower gate fidelities, whereas trapped-ion systems face challenges associated with scaling and motional-mode crowding in large ion crystals ~\cite{2-Liu_2025}.

The complementary advantages of neutral atoms and trapped ions motivate the development of hybrid platforms combining the strengths of both systems. Over the past two decades, such hybrid platforms have become an important tool for studying quantum interactions and many-body physics~\cite{11-RevModPhys.91.035001}. Numerous experiments have realized mixtures of trapped ions and ultracold neutral atoms~\cite{12-PhysRevA.95.032709,13-PhysRevLett.117.243401,14-PhysRevLett.102.223201,15-PhysRevA.96.030703}, revealing phenomena such as Feshbach~\cite{16-Weckesser_2021} and shape resonances~\cite{17-PhysRevA.55.636}. In these experiments, one or several ions are typically immersed in a cloud of ultracold atoms. Moreover, studies of atom–ion collision dynamics in Paul traps demonstrated the lower limits imposed by RF fields on the collision energy, and yet, observed quantum scattering effects far from the s-wave regime~\cite{13-PhysRevLett.117.243401,20-Pinkas_2023,21-PhysRevA.102.031301,22-Katz_2022,23-sikorsky2017spincontrolledatomioninelastic}. 

Alongside the experimental efforts, theoretical work has proposed a variety of hybrid Rydberg-atom–ion platforms, predicting novel interaction mechanisms and applications in quantum technologies. Among these are proposals for studying the interaction between a single Rydberg atom and a single trapped ion, which remain experimentally unexplored~\cite{58-PhysRevA.94.013420, 59-PhysRevLett.118.263201, 60-Wang_2020}. 

Here, we analyze a possible method to investigate this new regime of atom–ion interactions, involving a small number of neutral atoms individually trapped in optical tweezers and interacting with a few-ion crystal. In contrast to previous atom–ion experiments, that were based on atomic clouds, this approach enables deterministic control over both the number and position of the atoms relative to the ions. The tweezers geometry further provides precise control of the atom–ion separation, allowing access to the long-range interaction regime while reducing the influence of close-range Langevin collisions.

At large internuclear separations, the interaction between an ion and a neutral atom is dominated by the polarization potential arising from the electric field of the ion inducing a dipole moment in the atom. This interaction gives rise to an attractive long-range potential which scales as $V_{int} \sim -1/R^4$. Owing to its scaling, the polarization interaction extends over significantly larger distances than the van der Waals interaction between neutral atoms. As a result, it provides an efficient mechanism for coupling the motion and internal states of atoms and ions, even when they remain spatially separated.

By exciting the atom to a Rydberg state, its polarizability is significantly enhanced, increasing the strength of the polarization interaction. We find that such coupling can modify trapped ions’ collective motion and provides a mechanism to control quantum logic operations. Using this effect, we propose the implementation of a hybrid three-qubit controlled Mølmer–Sørensen gate, in which the internal state of the neutral atom determines the entangling interaction between two ions.

Overall, such a platform could enable controlled studies of atom–ion interactions at the single-particle level and may provide a route toward hybrid quantum interfaces combining neutral-atom and trapped-ion architectures.

\section{Theoretical Framework}

\subsection{Coupling internal atomic state to ionic motional state}
In an external electric field $\mathbf{E}$, a neutral atom acquires an induced dipole moment $\mathbf{p}=\mathbf{\alpha}\cdot\mathbf{E}$ characterized by the polarizability, $\alpha$. An important consequence of atomic polarizability is the resulting interaction between a charged particle and the dipole moment it induces on its neutral counterpart. This interaction leads to an attractive long-range polarization potential.\\
For Rydberg atoms, the acquired polarizability is exceptionally large, and scales approximately as $\sim n^7$, which gives rise to extreme sensitivity to external electric fields. Here we show that, since the atom interacts with the ions’ charge via its polarizability, the dynamics of the system strongly depend on the atom’s state.\\
For the theoretical description of the system, we begin by considering motion along the axial axis of the Paul trap, where only DC fields are present. The potential energy of the ionic crystal is modified when the atom is introduced into the system. Additional contributions arise from the interaction of the atom with the electric field generated by the ions and the Paul trap, which we denote as $V_{atom-DC field}$, as well as from the optical tweezer that confines the atom, denoted by $V_{atom-tweezers}$. The total potential energy of the system can therefore be written as,
\begin{equation}
    V=V_{ions}+V_{atom-DC field}+V_{atom-tweezers},
    \label{eq:total_potential}
\end{equation}
where:
\begin{equation}
    V_{ions}\equiv V_0 = \frac{1}{2} M_I \nu_I^2 (x_1^2+x_2^2)+\frac{Z^2e^2}{4\pi \epsilon_0} \frac{1}{|x_2-x_1|},
\end{equation}
\begin{equation}
    V_{atom-DC field} = -\frac{1}{2} \mathbf{p}\cdot\mathbf{E}_{DC}(x=x_A) = -\frac{1}{2} \alpha |\mathbf{E}_{DC}(x_A)|^2,
    \label{eq:atom_DC_potential}
\end{equation}
\begin{equation}
    V_{atom-tweezers} = \frac{1}{2}M_A\nu_A^2x_A^2.
    \label{eq:tweezers_trapping_potential}
\end{equation}

Here, $M_I$ and $M_A$ are the masses of an ion and the atom, respectively. The coordinates, $x_1$, $x_2$ and $x_A$ denote the positions of the two ions and the atom along the trap axis, and $\nu_I$ and $\nu_A$ are the trapping frequencies of the ions and the atom, respectively. The DC electric field at $x=x_A$ is given by the summation of the ions and the Paul trap fields,
\begin{equation}
    \label{eq:axial_DC_field}
    \mathbf{E}_{DC}(x_A)= \frac{Ze}{4\pi \epsilon_0}\left[ \frac{1}{|x_1-x_A|^2}-\frac{1}{|x_2-x_A|^2} \right] - \frac{M_I\nu_I^2}{e}x_A.
\end{equation}

For sufficiently large atomic polarizability, the presence of the atom modifies the normal-mode structure of the two-ion chain. In contrast, for small polarizability, the atom is effectively invisible to the ions, leaving their normal modes essentially unchanged.\\
To determine the normal modes of the new system, we expand the potential around the equilibrium positions and introduce a characteristic length scale l. For dimensionless normalized coordinates we use $u_i = \frac{x_{i}}{l}, \hspace{0.3cm} l=(\frac{Z^2 e^2}{4\pi \epsilon_0 M_I\nu_I^2}) ^{\frac{1}{3}}$, and the equilibrium positions of the ions and the atom are $u_{1,2}=\mp\left( \frac{1}{2} \right)^{\frac{2}{3}}, \hspace{0.3cm} u_A=0$ \cite{James_1998}.\\
To study small oscillations around the equilibrium, we write the coordinates of the ions and the atom as small displacements from equilibrium, $x_i=x_{0,i}+q_i$. The total potential can be expanded to second order, resulting quadratic Lagrangian describing coupled harmonic oscillators, 
\begin{equation}
L \approx 
\frac{M_I}{2}\left(\dot q_1^{2}+\dot q_2^{2}+ \dot q_A^{2}\right)
-\frac{1}{2}\sum_{i,j\in\{1,2,A\}}
C_iC_j
\frac{\partial^2V}{\partial x_i \partial x_j}
q_i q_j
\hspace{0.5cm}\equiv \hspace{0.5cm}
\frac{M_I}{2}\sum_i \dot{q_i}^2
- \frac{M_I\nu_I^2}{2}
\sum_{i,j\in\{1,2,A\}}
\hat A_{ij}\,q_i q_j.
\end{equation}

Here, $q_A=\sqrt{\frac{M_A}{M_I}}\,q_A'\equiv \frac{1}{\sqrt{r_m}}q'_A$ is the rescaled displacement of the atom, where $q_A'$ is the physical displacement and $r_m=\frac{M_I}{M_A}$ is the atom-to-ion mass ratio ; $C_{i=1,2}=1$, $C_a=\sqrt{r_m}$ are yet again rescaling constants ; $\left. \hat{A}_{ij} = \frac{1}{M_I\nu_I^2} C_iC_j\frac{\partial^2 V}{\partial x_i \partial x_j}\right|_0$ is the matrix of second derivatives of the potential at the point of equilibrium.\\
Evaluating the second derivatives of the potential at the equilibrium positions and diagonalizing the matrix $\hat{A}$, we obtain the mode structure vectors,

\begin{equation}
\label{eq:eigenvectors}
\vec v_1 = \frac{1}{\sqrt{2+\frac{a^2}{r_m}}}
\begin{pmatrix}
1\\
1\\
\frac{a}{\sqrt{r_m}}
\end{pmatrix},
\qquad
\vec v_2 = \frac{1}{\sqrt{2}}
\begin{pmatrix}
1\\
-1\\
0
\end{pmatrix},
\qquad
\vec v_3 = \frac{1}{\sqrt{2+\frac{b^2}{r_m}}}
\begin{pmatrix}
1\\
1\\
\frac{b}{\sqrt{r_m}}
\end{pmatrix},
\end{equation}

where,
\begin{equation}
\small
\begin{aligned}
a,b = \frac{1}{2^5(2^5+u^3)c_e\sqrt{r_m}} \Bigg[ c_e\left(2^9-(2^5+u^3)^2 r_m\right) +2^7 u^6(r_\nu^2-1) \pm \\
\sqrt{ c_e^2\left(2^9+(2^5+u^3)^2 r_m\right)^2 -2^8 c_e(-2^9+(2^5+u^3)^2 r_m)(-1+r_\nu^2)u^6 +2^{14}u^{12}(-1+r_\nu^2)^2 } \Bigg],
\end{aligned}
\end{equation}
and we denote: $u\equiv u_2-u_1$, $c_e\equiv \frac{2^7 \alpha M_I\nu_I^2}{e^2}$ and $r_\nu \equiv \frac{\nu_A}{\nu_I}$.\\
The corresponding eigenvalues are,
\begin{equation}
\label{eq:eigenvalues}
\small
\begin{aligned}
\lambda_{1,3} =
\frac{-c_e\left(2^9 + (2^5 + u^3)^2 r_m\right) + 2^7 u^6 (1 + r_\nu^2)}{2^8 u^6}
\pm \\
\frac{
\sqrt{
c_e^2\left(2^9 + (2^5 + u^3)^2 r_m\right)^2
- 2^8 c_e \left(-2^9 + (2^5 + u^3)^2 r_m\right)(-1 + r_\nu^2) u^6
+ 2^{14} u^{12} (-1 + r_\nu^2)^2
}
}{2^8 u^6}
\end{aligned},
\end{equation}
and,
\begin{equation}
\lambda_2 = 1 + \frac{4}{u^3}.
\end{equation}
In the following, we denote the normal modes' frequencies as $\nu_j=\sqrt{\lambda_j}$.\\
Since mode $\vec{v}_2$ represents symmetric motion of the ions around the atom equilibrium position, it remains decoupled from the motion of the atom, and therefore is not suitable  for atom-ion coupling. However, both the $\vec{v}_1$ mode and $\vec{v}_3$ can be used to this end. Here we focus our analysis on $\vec{v}_1$, which represents the modified center-of-mass (COM) mode of the two-ion crystal. For a two-ion crystal the COM frequency is $\nu_I$ with the corresponding eigenvalue is 1 \cite{James_1998}. The relative frequency shift of the COM mode in the presence of the atom can then be calculated to be,
\begin{equation}
    \label{eq:relative_shift}
    \frac{\nu_{1,with\hspace{1mm} atom}(\alpha)}{\nu_{1,no\hspace{1mm} atom}} \equiv \frac{\nu_1(\alpha)}{1}=\sqrt{\lambda_1}.
\end{equation}

\begin{figure}[htbp]
    \centering
    \includegraphics[width=\textwidth]{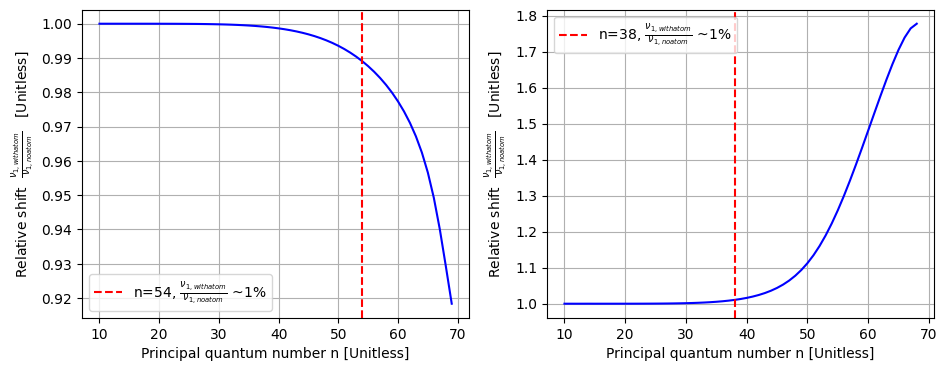}
    \caption{Relative shift of the center-of-mass (COM) motional mode frequency of a two-ion crystal induced by a nearby neutral $^{88}\mathrm{Sr}$ atom excited to Rydberg states. The shift is shown as a function of the principal quantum number $n$ for the $5sns\,({}^3S_1)$ (left) and $5snd\,({}^3D_1)$ (right) series. As $n$ increases, the atomic polarizability grows, leading to a stronger modification of the ion crystal’s motional frequency.}
    \label{fig:COM_frequency_shift}
\end{figure}

As we can see, the participation of the atom in the motion strongly depends on its polarizability. For a neutral atom in the ground state and low-lying n states $\alpha \rightarrow 0$, resulting $a\rightarrow 0$, $b\rightarrow \infty$, and consequently $\vec{v}_1\rightarrow (1,1,0)$ and $\vec{v}_3\rightarrow (0,0,1)$. This behavior is expected, since in this limit the motion of the atom becomes completely decoupled from the motion of the ions.\\
When the atom is excited to Rydberg states, its polarizability increases dramatically. In our calculations, the static polarizability was extracted by calculating the Stark map of a state, using the PairInteraction Python package \cite{56-Weber2017}, and fitting the electric field energy shift to the quadratic Stark relation, $\Delta E = -\frac{1}{2}\alpha E^2$. The fit was restricted to the low-field regime, in which the Stark shift remains quadratic and higher-order contributions due to state mixing are negligible.\\
Using the calculated static polarizabilities of the relevant states, the relative frequency shift was evaluated according to Eq.~\ref{eq:relative_shift}. In these calculations, the trapping frequencies of the ions and the atom were set to $\nu_I=2\pi \cdot 1MHz$, $\nu_A=2\pi \cdot 300kHz$, respectively. Fig.~\ref{fig:COM_frequency_shift} illustrates the relative shift of the COM mode frequency as a neutral $^{88}Sr$ atom is excited to higher n-states. The two panels show examples for the $5sns(^3S_1)$ and $5snd(^3D_1)$ Rydberg series. The former is characterized by a positive polarizability, whereas the latter exhibits a negative polarizability within the range of \textit{n} considered here. As \textit{n} increases, the magnitude of polarizability grows rapidly, resulting in an increasing shift of the COM mode frequency.

An additional observation arising from Eq.~\ref{eq:eigenvalues} is that the frequency shift depends strongly on the ions’ trapping frequency $\nu_I$. However, it is sensitive to the atom’s trapping frequency $\nu_A$ only when it is comparable to $\nu_I$. Therefore, from a practical point of view, the motional frequency shift is independent of the atom trapping frequency, which allows for flexibility in choosing the tweezers power.


\subsection{State-dependent control of the Mølmer–Sørensen gate}
In trapped-ion systems, laser fields couple the ions' internal states to their collective motion. Within the Lamb-Dicke regime, a laser tuned near a motional sideband drives transitions that simultaneously change the ion spin state and the phonon occupation of a collective vibrational mode \cite{wineland1998experimentalissuescoherentquantumstate}. This spin-motion coupling forms the basis for entangling gates, such as the M{\o}lmer--S{\o}rensen (MS) gate \cite{PhysRevA.62.022311}. In the MS gate scheme, the ions are illuminated by a bichromatic laser field consisting of two frequency components tuned near the red and blue motional sidebands, $\omega_{r,b}=\omega_{eg}\pm(\nu_I+\xi)$, where $\omega_{eg}$ is the carrier transition frequency, $\nu_I$ is the ions' motional frequency, and  $\xi$ is a small symmetric detuning from the sidebands. By choosing the gate parameters appropriately,
\begin{equation}
    \xi = 2\sqrt{n}\,\eta \Omega, \hspace{2cm} T = \frac{2\pi n}{\xi},
\end{equation}
The motional state undergoes a closed trajectory in phase space, returning to its initial state at gate time T, while the ions acquire a geometric phase that entangles their internal states, given by \cite{PhysRevA.62.022311, Shapira_2018},
\begin{equation}
    A(t)=\frac{\eta^2\Omega^2}{\xi} \left( t - \frac{sin(2\xi t)}{2\xi}\right).
\end{equation}
Here, $\Omega$ is the driving field Rabi frequency, $\eta$ is the Lamb-Dicke parameter, and n is the number of closed loops traversed by the motional state in phase space during the gate.\\
If the ions’ motional frequency is shifted, $\nu_I' = \nu_I + \Delta \nu_I$, the positions of the red and blue sidebands are correspondingly modified. This effect can be incorporated by introducing an effective symmetric detuning $\xi' = \xi - \Delta \nu_I$
such that the frequency shift can be interpreted as an additional contribution to the detuning.
If all other gate parameters are kept fixed, the evolution operator retains the same structure as in the standard MS gate, but the accumulated geometric phase gained by the ions is modified. At the gate time $T=\frac{2\pi n}{\xi}$, the resulting phase becomes
\begin{equation}
\begin{aligned}
    A'(t=T) = \frac{\eta^2\Omega^2}{\xi'} \left(T-\frac{sin(2\xi'T)}{2\xi'}\right) 
    = \frac{\eta^2\Omega^2}{\xi'} \left(\frac{2n\pi}{\xi}-\frac{sin(\frac{4\xi'n\pi}{\xi})}{2\xi'}\right) \\
    =\frac{\eta^2\Omega^2}{\xi^2r_\xi} \left( 2n\pi-\frac{sin(4\pi n r_\xi)}{4n r_\xi}\right)
    = \frac{1}{2 r_\xi}\left(\pi - \frac{\sin(4\pi n r_\xi)}{4 n r_\xi}\right)
\end{aligned}
\end{equation}
where we defined $r_\xi=\frac{\xi'}{\xi}$. 
Generally, the shifted gate results in a spin state that is not disentangled from the motion. But, choosing a value of $r_\xi$ that satisfies $n r_\xi \in \mathbb{Z}, \space n \in \mathbb{N} $, the spin–motion entanglement can still vanish at the end of the gate, while the accumulated phase at the end of the gate will be,
\begin{equation}
    A’(t=T) = \frac{\pi}{2r_\xi}.
\end{equation}
This mechanism allows the MS gate to become conditional on the motional frequency, enabling gate operations that depend on the presence of an additional particle or external perturbation that modifies the mode frequency.

\begin{figure}[htbp]
    \centering
    \includegraphics[width=\textwidth]{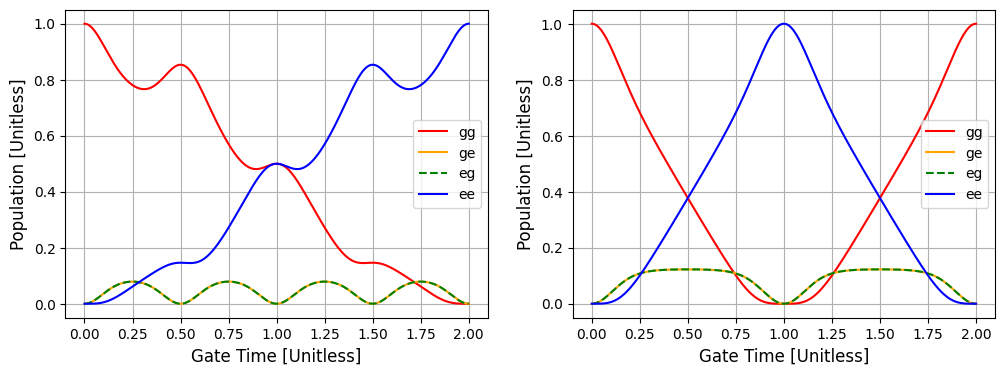}
    \caption{Time evolution of the two-ion state populations during a Mølmer--Sørensen (MS) gate. The left panel shows the standard gate dynamics when the motional frequency is unchanged, leading to the generation of an entangled state with equal populations in $|gg\rangle$ and $|ee\rangle$ at the gate time. The right panel illustrates the case where the motional frequency is shifted, modifying the accumulated phase and preventing the formation of the entangled state.}
    \label{fig:MS_gates}
\end{figure}

Figure~\ref{fig:MS_gates} shows the time evolution of the populations during a M{\o}lmer--S{\o}rensen (MS) gate for two ions in two scenarios: an unshifted gate and a gate with a shifted motional frequency. In the unshifted case ($n=2$, $r_\xi=1$), the populations evolve such that at the gate time $T=\frac{2\pi n}{\xi}$ the system reaches an entangled state with equal populations in $|gg\rangle$ and $|ee\rangle$. In contrast, when the motional mode is shifted so that the detuning ratio becomes $r_\xi=\tfrac{1}{2}$ (corresponding to a frequency shift $\Delta\nu_I=\xi/2$), the accumulated phase changes and the gate no longer produces entanglement at the same time $T$. Instead, the system returns close to a separable state, demonstrating how a controlled shift of the motional frequency can modify or suppress the MS gate operation.\\
In our proposed scheme, the presence of the neutral atom modifies the frequency of the vibrational mode (Fig.~\ref{fig:COM_frequency_shift}), thereby introducing an additional detuning to the MS gate. If the atom is in a low-lying state, with minimal polarizability, the gate proceeds essentially as usual. If the atom is in a Rydberg state, the induced shift in the mode frequency can alter or suppress the gate dynamics.\\
This mechanism enables a controlled three-body interaction: the state of the neutral atom determines whether the two-ion MS gate operates normally or is modified. In this way, the atom acts as a control qubit for an entangling operation between the ions. We have recently realized a similar scheme, in which a single ion, in a three-ion crystal, was held by a state-dependent tweezer potential. The phonon frequency of one of the crystal modes was therefore controlled in a state dependent way, leading to similar state-dependent control of an entanglement gate \cite{25-h4c6-463f}.

\section{Proposed Experimental Scheme}
\label{sec-experimental_system}


\subsection{Qubit encoding and trapping}
\begin{figure}[t!]
    \centering
    \includegraphics[width=1\textwidth]{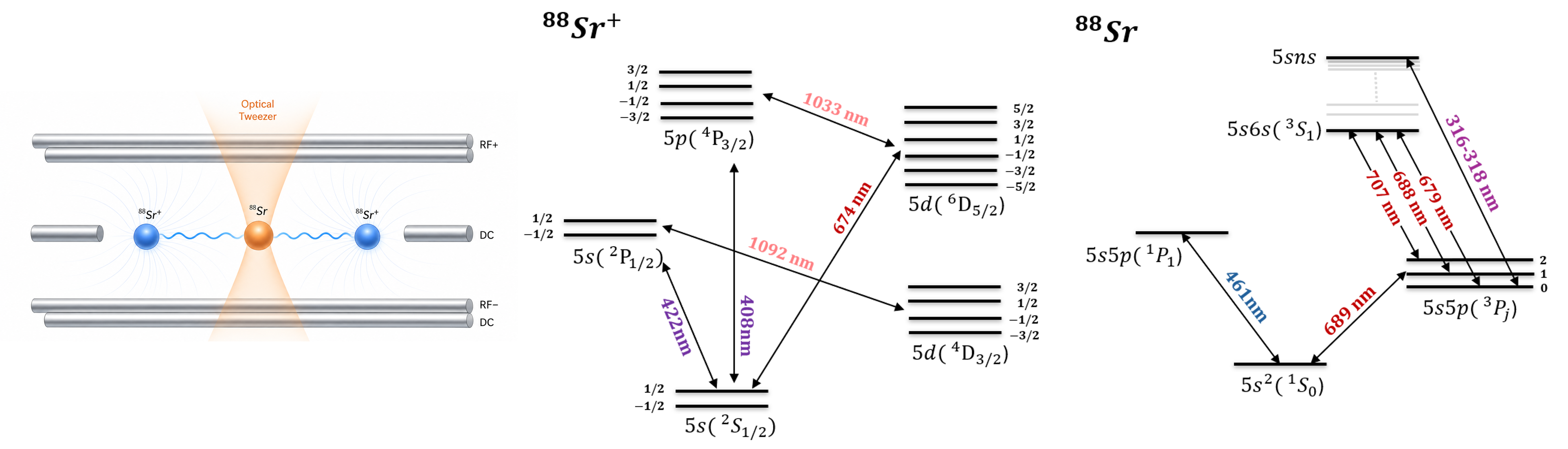}
    \caption{Left - proposed experimental system including two $^{88}\mathrm{Sr}^+$ ions confined in a linear Paul trap, while a single $^{88}\mathrm{Sr}$ atom is trapped at their midpoint using optical tweezers. For $^{88}\mathrm{Sr}^+$, the qubit can be encoded either in the optical transition between the $5s\,^2S_{1/2}$ and $4d\,^2D_{5/2}$ states or in the Zeeman sublevels of the ground state. For the neutral atom, the qubit is encoded between the metastable clock state $5s5p\,({}^3P_0)$ and a highly excited Rydberg state. Middle and right panels show the relevant energy levels and optical transitions for $^{88}\mathrm{Sr}^+$ and $^{88}\mathrm{Sr}$, respectively, including the transitions used for cooling, state preparation, and coherent qubit manipulation.}
    \label{fig:experimental_scheme}
\end{figure}

We consider a system consisting of two $^{88}\mathrm{Sr}^+$ ions, and a single $^{88}\mathrm{Sr}$ atom trapped at their midpoint. For the implementation of quantum operations in this scheme, qubits will be encoded in the internal energy levels of both the atom and the ions. For $^{88}\mathrm{Sr}^+$, the qubit can be encoded either in an optical transition between the $S_{1/2}$ and $D_{5/2}$ states, or in the Zeeman sublevels of the ground state. For the atom, the qubit will be encoded between the metastable clock state $5s5p\,({}^3P_0)$ and a highly excited Rydberg state. The meta-stable clock state provides a long lifetime, while the Rydberg state enables strong interactions via its large static polarizability. The choice of the meta-stable clock state as a qubit level, instead of the electronic ground state, has to do with its dynamic polarizability under the tweezers potential as detailed below. The proposed experimental system and the relevant atomic levels are illustrated in Fig.~\ref{fig:experimental_scheme}.

\begin{figure}[htbp]
    \centering
    \includegraphics[width=0.5\textwidth]{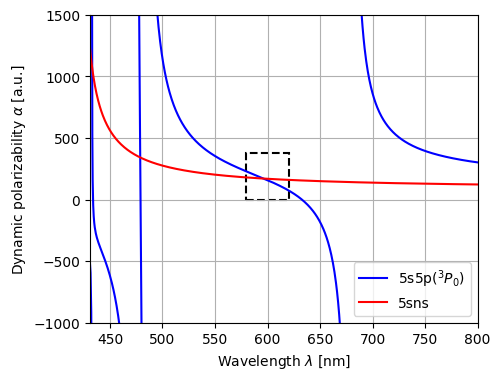}
    \caption{Dynamic polarizability of $^{88}\mathrm{Sr}$ as a function of wavelength, using  a simplified model. The blue and red curves show the polarizabilities of the $5s5p\,({}^3P_0)$ clock state and $5sns$ Rydberg states, respectively. The intersection near $596\,\mathrm{nm}$ indicates the magic wavelength where both states experience the same light shift, enabling state-insensitive optical trapping.}
    \label{fig:magic_trapping}
\end{figure}

\begin{figure}[t!]
    \centering
    \includegraphics[width=1.05\textwidth]{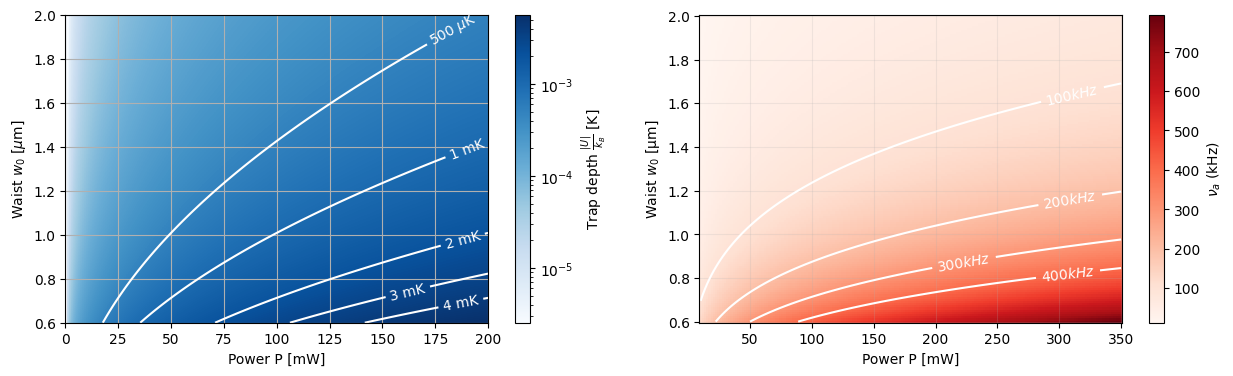}
    \caption{Calculated trapping parameters for optical tweezers at the magic wavelength of $596\,\mathrm{nm}$. The right panel shows the expected trapping frequency as a function of laser power and beam waist, while the left panel presents the corresponding trap depth under the same conditions.}
    \label{fig:trap_freq_depth}
\end{figure}

The ions are confined in a linear RF Paul trap using a combination of time-dependent and static electric fields.
Radial confinement is provided by an oscillating quadrupole potential of the form $\Phi(x,y,t) = U_{\mathrm{RF}} \cos(\Omega_{\mathrm{RF}} t)(x^2 - y^2)$, while static (DC) fields provide confinement along the axial direction. This configuration results in the formation of a linear Coulomb crystal with well-defined collective motional modes.\\
Neutral atoms are trapped using optical tweezers formed by tightly focused Gaussian laser beams, which create three-dimensional trapping potentials. For red-detuned light, atoms are attracted to the intensity maximum at the beam focus, leading to stable confinement. The strength of the confinement determines the trapping frequency of the tweezers.\\
The choice of atomic species arises from convenient trapping considerations, alongside the advantages of using the same atomic specie for both atoms and ions. A key requirement in our setup is for the atom to be stably trapped in both ground state and Rydberg state, as we would like to facilitate the interaction between the atoms and the ions. Furthermore, to avoid atomic dephasing caused by the tweezers, we would require that the optical trapping potential in both states will be identical. In the short-wavelength regime, the polarizability of Rydberg states is dominated by the ionic core, an effect that is negligible in alkali atoms but significant in multi-electron systems \cite{PhysRevA.88.043407}. In alkaline-earth atoms, positive polarizability contribution from the core can compensate the free-electron term \cite{PhysRevLett.128.033201, Topcu_2014}, enabling trapping of both ground and Rydberg states in conventional red-detuned dipole traps. For $^{88}$Sr specifically, a universal magic wavelength near 596~nm was found \cite{Topcu_2014}, for which the polarizabilities of the clock state and all Rydberg states higher than n=50 become equal and independent of n.\\
Fig.~\ref{fig:magic_trapping} presents the predictions of a simplified model of the dynamic polarizability of $^{88}$Sr, in which the contribution of the Rydberg electron is neglected, and only the ionic core contribution is considered. The data used in this analysis was extracted from Safronova's Portal for High-Precision Atomic Data and Computation \cite{55-2025Portal}.
We estimate the dynamic polarizability value at the magic wavelength to be approximately $\sim170$ [a.u.].\\ 
Using this value, we calculate the expected trapping frequency and trap depth for optical tweezers with varying laser power and beam waist. The analytical results are shown in Fig.~\ref{fig:trap_freq_depth}. As our scheme aims to cool the trapped atom to temperatures of $1\,\mathrm{mK}$ and below, the right panel of Fig.~\ref{fig:trap_freq_depth} shows that, for a realistic optical tweezers waist of approximately $1\,\mu\mathrm{m}$, optical powers of $100\,\mathrm{mW}$ or higher are required. However, as discussed in the following subsection, stability considerations may further constrain the range of suitable tweezers parameters, with the optimal operating conditions depending on the atomic state of interest.

\subsection{Stability of a Rydberg atom in the combined fields of the Paul trap, ions, and optical tweezers}

\begin{figure}[htbp]
    \centering
    \includegraphics[width=1\textwidth]{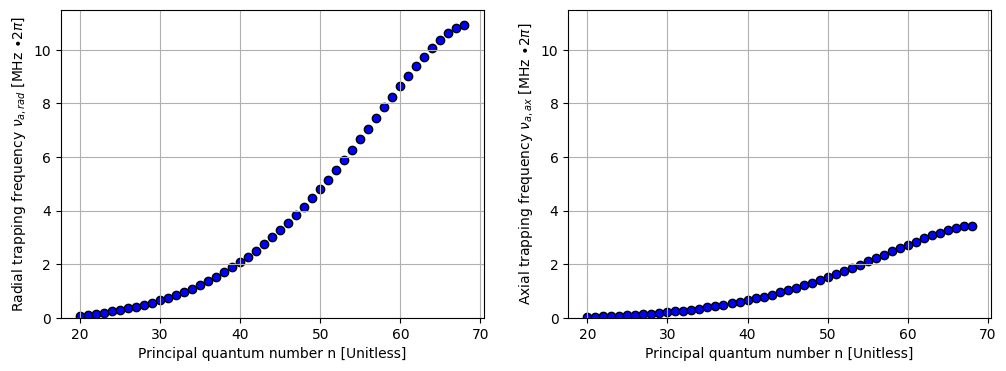}
    \caption{Radial (left) and axial (right) trapping frequencies of the $5snd(^3D_1)$ Rydberg state of Sr as a function of the principal quantum number, calculated from the total trapping potential, including both the optical tweezers and the ions, and the Paul trap potential. The considered Paul trap parameters are $V_{\rm RF}\approx200~\mathrm{V}$ and $r_0\approx0.27~\mathrm{mm}$.}
    \label{fig:Paul_trapping}
\end{figure}

We investigate the stability of the trapping potential experienced by the atom when it is excited to a Rydberg state, whose large static polarizability makes it particularly sensitive to external electric fields. Stable confinement requires that the total potential be locally convex around the equilibrium position,
\begin{equation}
    \frac{\partial^2 V_{\mathrm{tot}}}{\partial x_a^2}>0.
    \label{eq:stability_condition}
\end{equation}
This criterion ensures that the equilibrium corresponds to a local minimum of the potential and therefore that the atom remains trapped. The total potential includes contributions from the optical tweezers, the electric fields of the Paul trap, and the Coulomb fields generated by the trapped ions.\\
The effective trapping potential differs along the axial and radial directions of the Paul trap. Along the trap axis, the atom experiences the static DC electric field generated by both the Paul trap electrodes and the ions, together with the optical tweezers potential. In the radial direction, the atom is subjected to the oscillating RF quadrupole field of the Paul trap. Since the RF drive frequency $\Omega_{\mathrm{RF}}$ is much larger than the characteristic timescale of the atomic motion, the atom does not follow the rapid field oscillations and instead experiences the time-averaged potential of the RF field. Therefore, static polarizability should be taken into account. Expanding the total potential around the equilibrium position yields
\begin{equation}
    V_{axial} = -\frac{1}{2}\alpha |\mathbf{E}_{DC}(x_\Delta)|^2 + \frac{1}{2} M_a \nu_a^2 (x_\Delta)^2,
\end{equation}
\begin{equation}
    V_{radial}=
    \begin{cases}
    -\frac{1}{4}\alpha \left( \frac{V_RF}{r_0^2} \right)^2 (x_\Delta)^2 + \frac{1}{2}M_a \nu^2_{a,r} (x_\Delta)^2\\
    -\frac{1}{4}\alpha \left( \frac{V_RF}{r_0^2} \right)^2 (x_\Delta)^2 + \frac{1}{2}M_a \nu^2_{a,z} (x_\Delta)^2
\end{cases}
,
\end{equation}
where $x_\Delta$ denotes the displacement of the atom from its equilibrium position, $\nu_{a,r}$ and $\nu_{a,z}$ are the radial and axial trapping frequencies of the optical tweezers, respectively, and $\mathbf{E}_{\mathrm{DC}}$ is given by Eq.~\ref{eq:axial_DC_field}. Note that the optical tweezers provide anisotropic confinement, with a much stronger intensity gradient in the transverse directions than along the beam propagation axis \cite{61-grimm1999opticaldipoletrapsneutral}. Consequently, the radial trapping frequency is typically much larger than the axial one. In the following calculations, we neglect this anisotropy and use only the radial trapping frequency.\\
The sign of the Rydberg-state polarizability determines whether the electric fields of the Paul trap enhance or oppose the optical confinement. For states with negative polarizability, the atom is low-field seeking and is trapped near the RF null. In this case, both the RF potential and the optical tweezers contribute to confinement, so the Paul trap does not impose any additional stability constraint. In contrast, states with positive polarizability are high-field seeking and are attracted toward regions of large electric field. The RF potential therefore becomes anti-trapping, and stable confinement relies on the restoring force provided by the optical tweezers. As a result, a minimum optical trapping frequency is required to satisfy Eq.~(\ref{eq:stability_condition}). For representative experimental parameters, $V_{\mathrm{RF}}\approx200~\mathrm{V}$ and $r_0\approx0.27~\mathrm{mm}$, the RF-induced anti-trapping is sufficiently strong, indicating that extremely strong optical confinement is required to maintain stable trapping of the atom.\\
These considerations make Rydberg states with negative polarizability, such as the $5snd(^3D_1)$, considerably more practical for the proposed hybrid system. Figure~\ref{fig:Paul_trapping} shows the resulting radial and axial trapping frequencies as a function of the principal quantum number, obtained from the full trapping potential including the optical tweezers, the Paul trap, and the trapped ions. The figure shows that the confinement provided by the Paul trap becomes increasingly strong with principal quantum number, leading to rapidly increasing trapping frequencies for negative-polarizability Rydberg states.

Interestingly, for sufficiently high principal quantum numbers, the trapping frequency exceeds that of an ion confined in a Paul trap with the same trap parameters. For the parameters considered here, a singly charged ion would have a trapping frequency of approximately $\nu_I = 1~\mathrm{MHz}$. This seemingly counterintuitive result can be explained by the different field-particle interactions. The ion couples linearly to the oscillating RF electric field, resulting in a pseudopotential, whereas the Rydberg atom experiences the quadratic Stark potential, $U=-\frac{1}{2}\alpha E^2$, and therefore couples directly to the time-averaged electric-field intensity. Since the Rydberg-state polarizability increases rapidly with the principal quantum number, the resulting trapping frequency can exceed the ion secular frequency for sufficiently large $n$.


\subsection{Relaxation Times - $T_1$, $T_2$}
\begin{figure}[htbp]
    \centering
    \includegraphics[width=0.55\textwidth]{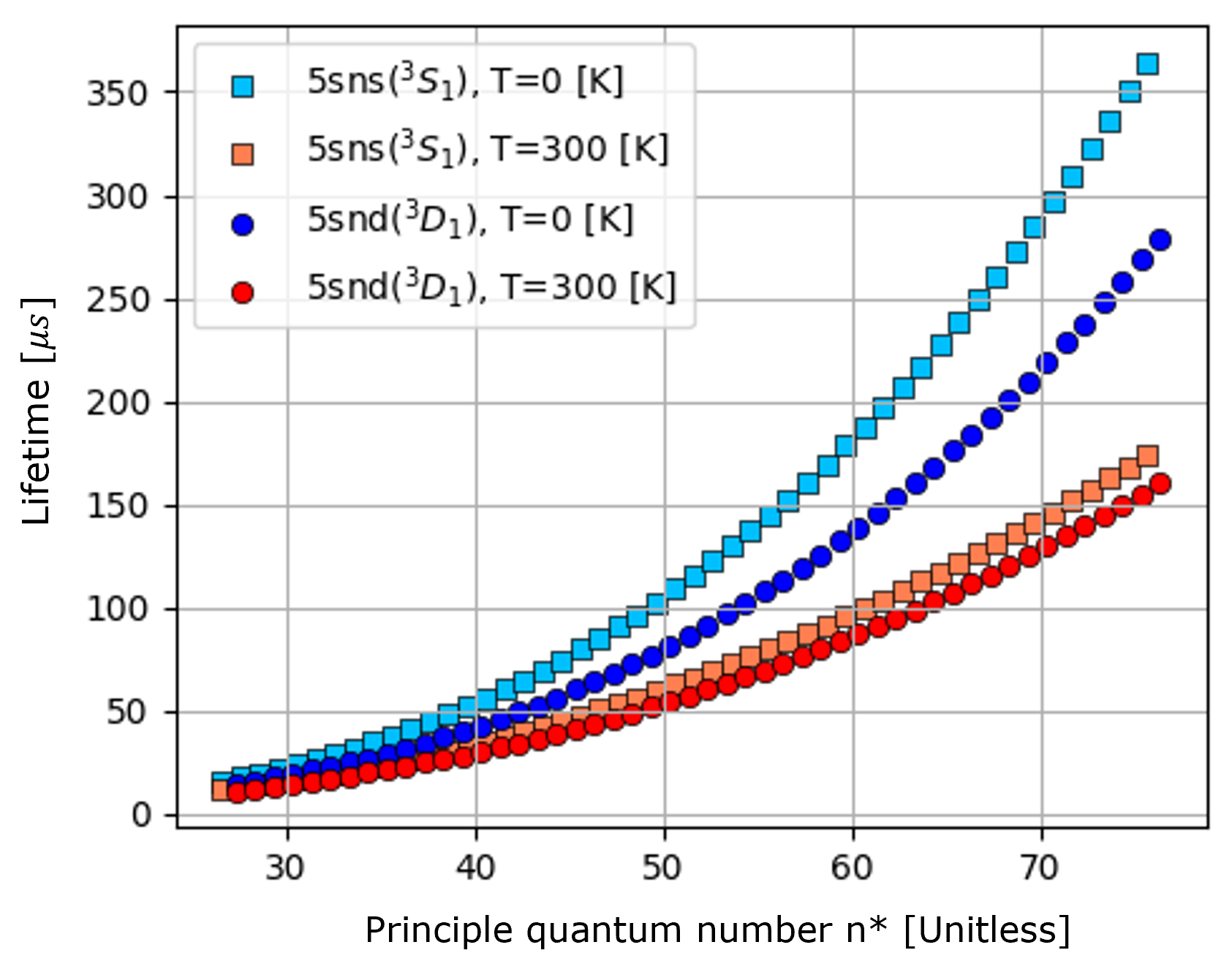}
    \caption{Calculated lifetimes of the $5sns(^3S_1)$ and $5snd(^3D_1)$ Rydberg series of $^{88}$Sr as a function of the effective principal quantum number $n^*$, for temperatures of $T=0$~K and $T=300$~K.}
    \label{fig:Sr_lifetimes}
\end{figure}

Since our key goal is an entanglement gate operation on the ions, controlled by the atom’s internal state, the lifetime of the Rydberg state must exceed the timescale set by the atom–ion interaction, requiring a lifetime of $\sim 100\,\mu s$ for mode shifts of a few 10’s kHz. A long Rydberg-state lifetime is therefore essential for achieving our goal.\\
In a room-temperature environment, Rydberg atoms are strongly affected by blackbody radiation (BBR). This is primarily because the small energy spacing between neighboring Rydberg levels overlaps with the thermal radiation spectrum at 300 K, resulting in rapid BBR-induced transitions among nearby states \cite{52-TF_Gallagher_1988}. At cryogenic temperatures, the BBR spectrum is strongly suppressed, and the lifetime of an atom in a Rydberg state is instead determined primarily by the spontaneous decay rate to lower-lying states. Owing to the scaling of the transition matrix elements and transition frequencies with the principal quantum number, the total spontaneous decay rate decreases for highly excited Rydberg states, resulting in their characteristically long lifetimes.\\
Fig~\ref{fig:Sr_lifetimes} presents the estimated lifetimes of $^{88}Sr$ Rydberg states for various principal quantum numbers. The calculations were carried out in Python using the PairInteraction package \cite{56-Weber2017}, and implements single-channel quantum defect theory (SQDT), which quantifies the deviations of many-electron atoms from hydrogenic energy levels \cite{51-MJ_Seaton_1966, 57-ROBERTSON2021107814}. At cryogenic temperatures, the calculated lifetimes exceed $100~\mu\mathrm{s}$ for the $5sns(^3S_1)$ series when $n \gtrsim 50$. At room temperature, comparable lifetimes are reached only for $n \gtrsim 60$. The $5snd(^3D_1)$ series exhibits shorter lifetimes over the entire range of principal quantum numbers considered, although its lifetime also exceeds $100~\mu\mathrm{s}$ for $n \gtrsim 65$ at room temperature, and $n \gtrsim 55$ at cryogenic temperatures.

The estimates presented above account only for population decay processes and therefore correspond to the Rydberg-state lifetime $T_1$. In realistic experiments, the coherence time $T_2$ may be further limited by dephasing mechanisms such as fluctuating Stark shifts induced by stray electric fields, for example, due to charge accumulation on nearby dielectric surfaces; laser-induced light shifts; spatial inhomogeneity of the excitation fields; Doppler shifts; laser amplitude and phase noise; blackbody-radiation-induced Stark shifts.\\
As discussed previously, we choose to operate the optical tweezers at the magic wavelength in order to suppress dephasing arising from fluctuations in the trapping laser intensity. We expect that another significant contribution to the reduction of the coherence time, $T_2$, will originate from instabilities of the Paul trap RF field. Since the ground and Rydberg states experience different RF trapping potentials, fluctuations in the RF field lead to fluctuations in the transition frequency between these states, resulting in dephasing.\\
To estimate the required stability of the RF trapping field we analyze how RF-induced fluctuations in the motional frequency effect the atomic transition. Since the clock and Rydberg states experience different trapping potentials, the optical transition frequency can be written as
\begin{equation}
\omega = \omega_0 + \frac{\hbar}{2}\left( \omega_{\mathrm{motional}}^{(\mathrm{clock})} - \omega_{\mathrm{motional}}^{(\mathrm{Rydberg})} \right) = \omega_0 + \Delta\omega_{\mathrm{motional}} + \delta(t),
\end{equation}
where $\Delta\omega_{\mathrm{motional}}$ is the static shift arising from the different motional frequencies associated with the clock and Rydberg states, while $\delta(t)$ represents temporal fluctuations caused by noise in the Paul trap RF drive. The static contribution can be calibrated and tracked experimentally, whereas the fluctuating term cannot be compensated and therefore leads to dephasing.\\
Assuming zero-mean fluctuations with variance $\langle \delta^2 \rangle$, we require the accumulated phase uncertainty during the gate time, $\Delta\delta\,T$, to remain much smaller than $\pi$. For a gate duration of $T=100~\mu\mathrm{s}$, this yields the condition $\Delta\delta \ll 3.1\times10^4~\mathrm{Hz}$. Then, the required relative RF stability is
\begin{equation}
\frac{\Delta\Omega_{\mathrm{RF}}}{\Omega_{\mathrm{RF}}} \ll \frac{3.1\times10^4}{22\times10^6} \simeq 1.4\times10^{-3},
\end{equation}
where $\Omega_{\mathrm{RF}} = 22~\mathrm{MHz}$ is the RF drive frequency used in our trap.


\section{Discussion and Conclusions}
\label{sec-discussion}

In this work, we analyzed a hybrid atom-ion platform consisting of individually trapped neutral atoms interacting with a small trapped-ion crystal. We showed that the large polarizability of Rydberg atoms can induce measurable shifts in the collective motional modes of the ions, enabling the atomic internal state to influence ion-based quantum gate operations. In particular, we showed that this mechanism can modify the accumulated phase during a M{\o}lmer--S{\o}rensen gate, thereby enabling a conditional interaction in which the state of the atom controls an entangling operation between two ions.

Using a simplified theoretical model, we investigated the dependence of the motional mode structure on the atomic polarizability and estimated the experimental requirements for observing this effect. In particular, we analyzed the trapping conditions, optimal polarizabilities, Rydberg-state lifetimes, and dephasing mechanisms relevant for realistic implementations. 

Our results suggest that optical tweezers control of Rydberg atoms combined with trapped-ion quantum logic, offers a promising route toward controllable hybrid quantum systems with single-particle resolution. More generally, the ability to engineer state-dependent modifications of ion-crystal motional modes may enable new approaches for hybrid quantum gates, quantum simulation, and studies of many-body dynamics in mixed atom-ion systems. During the preparation of this manuscript, we became aware of a related proposal by Mudli ~\cite{54-Mudli_2026}, which considers an ion-atom two-qubit quantum gate based on phonon blockade.


\newpage

\appendix
\centerline{\Large \bf Appendix}

\section{Validity of the dipole approximation}
\label{sec-Validity}

Throughout this work, the interaction between the Rydberg atom and the electric fields generated by the ions and the Paul trap is treated perturbatively, retaining terms up to second order in the electric field. This approximation is valid provided two conditions: the interaction energy remains much smaller than the spacing between neighboring Rydberg levels ; The characteristic length scale of the Rydberg electron orbit is much smaller than the atom-ion distance. We therefore define the ratio between the interaction energy ($V_{atom-field}$) and the Rydberg energy level ($E_n$) spacing,
\begin{equation}
\label{eq:validity_dipole_approx}
r_E \equiv \frac{\left|V_{\mathrm{atom-field}}\right|}
{E_{n+1}-E_n},
\end{equation}
and the ratio between the characteristic Rydberg electron orbit ($R_{Ryd}$) and the atom-ion spacing ($d_{atom-ion}$),
\begin{equation}
    r_r \equiv\frac{r_{Ryd}}{d_{atom-ion}} \sim \frac{1}{d}\frac{4\pi\epsilon_0 \hbar^2 n_{eff}^2}{e^2 m_e},
\end{equation}
and require $r_E,r_r \ll 1$.

\begin{figure}[htbp]
    \centering
    \includegraphics[width=0.52\textwidth]{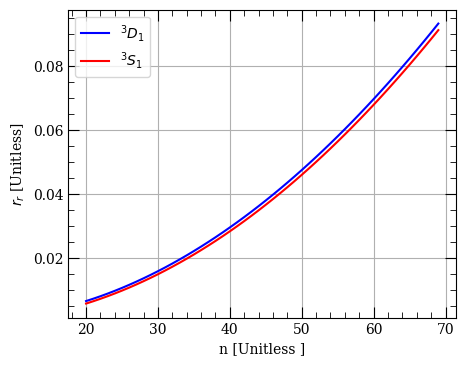}
    \caption{Ratio between the estimated Rydberg radius $r_{\mathrm{Ry}}$ and the atom-ion separation $d_{\mathrm{atom-ion}}$ as a function of the principal quantum number $n$ for the $5snd(^3D_1)$ (blue) and $5sns(^3S_1)$ (red) states. The ratio remains below $0.1$ over the entire range considered, indicating that the spatial extent of the Rydberg orbital is significantly smaller than the atom--ion separation.}
    \label{fig:ratio_rr}
\end{figure}

Note that we estimate the atomic radius using the simple Bohr model \cite{Rydberg_atoms}, as any atom at the Rydberg level can be reasonably approximated as a hydrogen atom, up to quantum defect corrections. The dependence of $r_r$ on the principal quantum number $n$ is shown in Fig.~\ref{fig:ratio_rr} for the $^5snd(^3D_1)$ and $5sns(^3S_1)$ Rydberg states. Throughout the range considered, the characteristic size of the Rydberg orbital remains significantly smaller than the atom--ion separation, supporting the validity of the point-like dipole approximation.

For a Rydberg state with principal quantum number $n$, the energy levels are given by the Rydberg–Ritz formula, $E_n \simeq -\frac{\mathrm{Ry}_{\infty}}{(n-\delta)^2}$, where $\delta$ is the quantum defect, and $Ry_\infty$ is the Rydberg constant.\\
As seen in Eq.~\ref{eq:total_potential} The atom-field interaction energy is given by the quadratic Stark shift, $V_{\mathrm{atom-field}} = -\frac{1}{2}\alpha |\mathbf{E}|^2$, where $\alpha$ is the static polarizability of the Rydberg state and $\mathbf{E}$ is the electric field described by Eq.~\ref{eq:axial_DC_field} for the axial direction, and described by the Paul trap's RF field $\textbf{E}_{RF}=-\frac{V_{RF}}{r_0^2}cos(\Omega t)(x\hat{x}-y\hat{y})$ for the radial direction.\\
Since the validity criterion Eq.~\ref{eq:validity_dipole_approx} becomes most restrictive at the largest electric field experienced by the atom, it is sufficient to evaluate it at the maximum field. We therefore estimate the electric field when the COM mode reaches its maximum displacement amplitude A. Using the relations given by the COM eigenvector $\vec{v}_1$ (Eq.~\ref{eq:eigenvectors}) and the DC electric field (Eq.~\ref{eq:axial_DC_field}), we obtain the maximum electric field in the axial direction,
\begin{equation}
    E_{axial,max} \approx \frac{Z^2e}{4\pi\epsilon_0 l^2} 
    \left[ \frac{1}{|-\frac{u}{2}+(1-a)\frac{A}{l}|^2}-\frac{1}{\frac{u}{2}+(1-a)\frac{A}{l}|^2} \right] 
    - \frac{aA}{l},
\end{equation}
where l is the length scale, and u is the equilibrium position of the ions in the normalized coordinates. Using the expression for Paul trap's RF field, averaging it in time, we get the maximum electric field in the radial direction,
\begin{equation}
    E_{radial,max} \approx -\frac{1}{4} \alpha \left( \frac{V_{RF}}{r_0^2} A^2 \right).
\end{equation}

\begin{figure}[b!]
    \centering
    \includegraphics[width=0.9\textwidth]{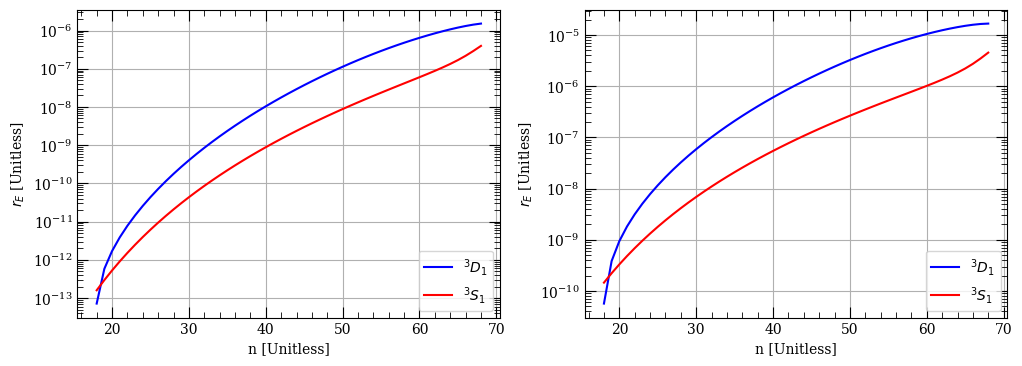}
    \caption{The ratio $r_E$ as a function of the principal quantum number $n$ for the $S_{1}$ (red) and $D_{1}$ (blue) Rydberg states. The left and right panels correspond to displacement in the axial and radial directions, respectively. Although $r_E$ increases with $n$ due to the growing polarizability of the Rydberg states, it remains well below unity over the entire range considered, validating the induced-dipole approximation.}
    \label{fig:ratio_rE}
\end{figure}

The amplitude A is determined by the atom's temperature, which is equivalent to the temperature of the motional mode. The energy of a quantum harmonic oscillator at thermal equilibrium is given by the Boltzmann distribution, $E=\hbar\nu \left( \frac{1}{e^{\frac{\hbar \nu}{k_B T}}-1}+\frac{1}{2} \right)$, leading us to the expression for the classical amplitude of motion:

\begin{equation}
    A\approx \sqrt{\frac{2\hbar}{m\nu} \left( \frac{1}{e^{\frac{\hbar \nu}{k_B T}}-1}+\frac{1}{2} \right)}.
\end{equation}
For an upper bound temperature $T=1 mK$, we'll get $A\approx0.07 \mu m=0.02l$.

We calculate the ratio $r_E$ (Eq.~\ref{eq:validity_dipole_approx}) as a function of the principal quantum number $n$. Figure~\ref{fig:ratio_rE} shows the resulting ratio for both the $S_{1}$ and $D_{1}$ states, considering the axial and radial modes. In all cases, $r_E \ll 1$, indicating that the interaction energy remains much smaller than the spacing between neighboring Rydberg levels throughout the parameter range considered.

\section{Ionization of Rydberg atoms by the DC and RF fields}
\label{sec-Ionization}
An additional limitation when working with Rydberg atoms is the possibility of field ionization. As the principal quantum number n increases, the binding energy of the outer electron decreases and the energy gap to the ionization continuum becomes smaller, making the atom more susceptible to ionization. In particular, external electric fields—such as those generated by the ions or the trapping potentials—can increase the tunneling probability of the excited electron. Approximating the Rydberg atom as a hydrogen-like system, the tunneling ionization rate in a DC electric field E can be estimated as:
\begin{equation}
    \gamma_{ionization}
    =
    16 \frac{\varepsilon_{ion}^{2}}{\hbar e |E| r_n}
    \exp\!\left[
    -\frac{4}{3}\frac{\varepsilon_{ion}}{e|E|r_n}
    \right],
    \qquad
    \varepsilon_{ion}=\frac{Ry_\infty}{(n-\delta_{def})^2},
    \qquad
    r_n=\frac{4\pi\epsilon_0 \hbar^2 n^2}{e^2 m_e}
\end{equation}
where $\varepsilon_{ion}$ is the ionization energy, $Ry_\infty$ is the Rydberg constant, $\delta_{def}$ is the quantum defect, and $a_0$ is the Bohr radius.

\begin{figure}[htbp]
    \centering
    \includegraphics[width=1\textwidth]{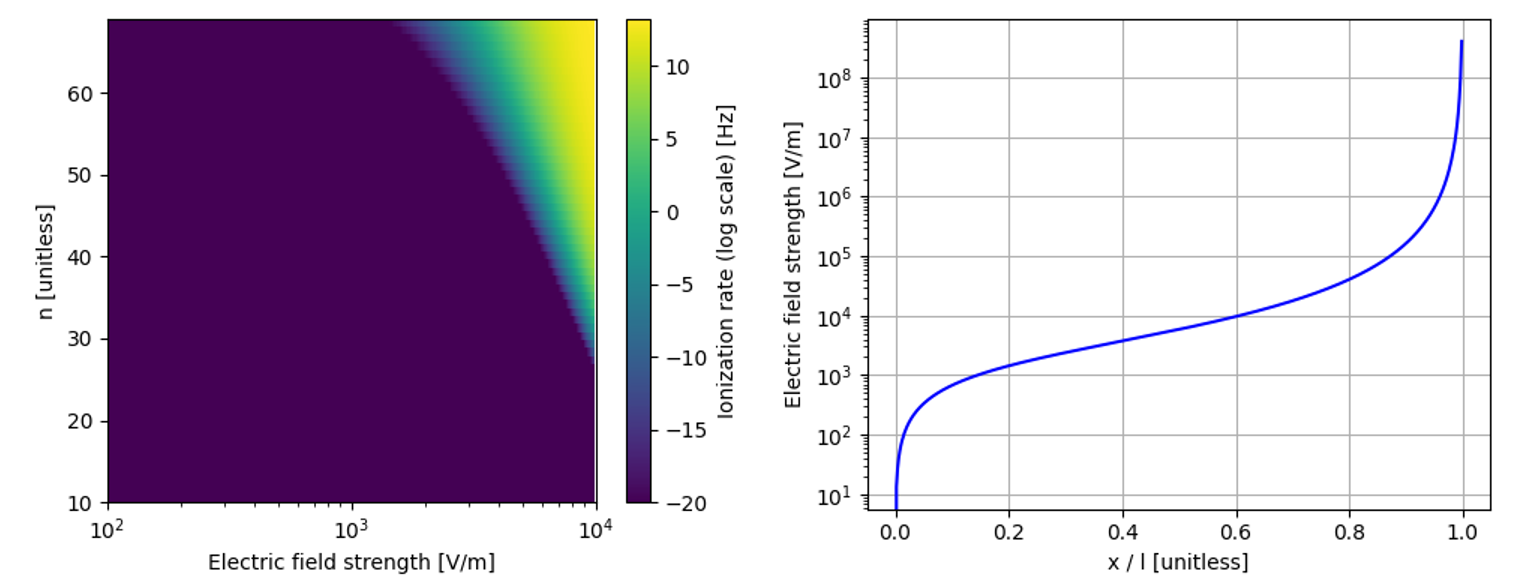}
    \caption{electric-field ionization of $^{88}\mathrm{Sr}$ Rydberg atoms. The left panel shows the calculated ionization rate as a function of the principal quantum number $n$ and the applied electric field strength, illustrating the strong exponential increase of the ionization probability for higher Rydberg states and stronger fields. The right panel shows the electric field generated by two ions as a function of the atom’s position between them.}
    \label{fig:ionization_rate}
\end{figure}

Figure~\ref{fig:ionization_rate} presents the calculated field-ionization rate of a Rydberg atom as a function of the principal quantum number n and the electric field strength. The color map shows the ionization rate on a logarithmic scale, illustrating the strong exponential dependence of the tunneling ionization process on the applied electric field. As expected, higher Rydberg states (larger n) are significantly more susceptible to ionization due to their reduced binding energy. The second panel shows the magnitude of the electric field generated by the DC-fields as a function of the atom’s position between the ions. The field grows rapidly as the atom approaches the ion, while remaining relatively weak near the center of the configuration.\\
As for the oscillating RF field of the Paul trap, for typical operating parameters of our trap ($V_RF\approx200V$, $r_0\approx 0.27mm$) the RF field amplitude near the trap center remains well below the threshold required for Rydberg field ionization. Moreover, the RF field vanishes at the trap center (the RF null), where the atom is positioned in our scheme, and increases only linearly with displacement. Consequently, for the small motional amplitudes considered here, the RF field does not significantly contribute to the ionization probability.

\section{The anharmonic effect of temperature on the modes}
\label{sec-Anharmonic}

The normal-mode analysis presented in the main text is based on the harmonic approximation, obtained by expanding the total potential to second order around the equilibrium positions. Although this approximation is mostly justified, at finite temperature, the particles oscillate with a finite amplitude, making the higher-order, anharmonic terms of the potential non-negligible. As a result, the normal-mode frequencies acquire a temperature-dependent correction. We therefore would like to assess the effect of non-zero temperature on the COM frequency.\\
To estimate this effect, we numerically solved the full nonlinear equations of motion derived from the complete Lagrangian, without invoking the harmonic approximation,
\begin{align}
L &= \frac{1}{2}M_I\left(\dot{x}_1^2+\dot{x}_2^2\right)
+\frac{1}{2}M_a\dot{x}_a^2
-V(x_1,x_2,x_a),
\end{align}

where
\begin{align}
V(x_1,x_2,x_a)
&=
\frac{1}{2}M_I\nu_I^2\left(x_1^2+x_2^2\right)
+\frac{1}{2}M_a\nu_a^2x_a^2
+\frac{Z^2e^2}{4\pi\epsilon_0}\frac{1}{|x_2-x_1|}
\nonumber\\
&\quad
-\frac{1}{2}\alpha
\left(\frac{Z^2e}{4\pi\epsilon_0}\right)^2
\left[
\frac{1}{|x_2-x_a|^2}
-\frac{1}{|x_1-x_a|^2}
-\frac{x_a}{l^3}
\right]^2,
\end{align}

and the equations of motion are
\begin{align}
\frac{d}{dt}\frac{\partial L}{\partial\dot{\vec{x}}}
=
\frac{\partial L}{\partial\vec{x}}.
\end{align}

\begin{figure}[htbp]
    \centering
    \includegraphics[width=0.9\linewidth]{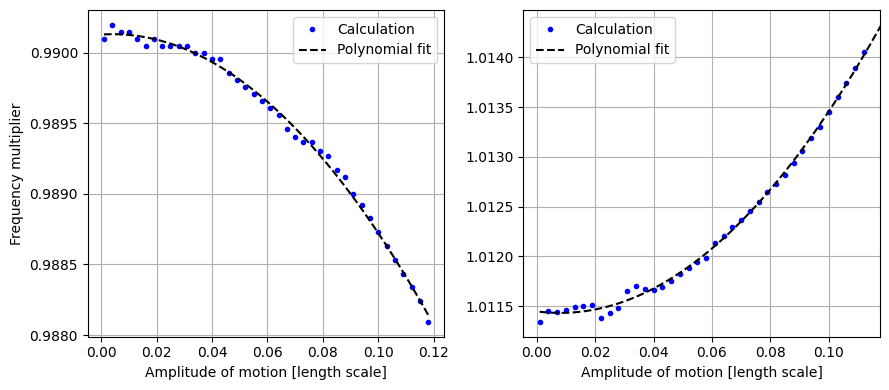}
    \caption{COM frequency multiplier obtained from numerical integration of the full equations of motion as a function of the oscillation amplitude (normalized by the $l$). The left and right panels correspond to atomic polarizabilities of $\alpha=+70\,\hbar\,\mathrm{MHz}\,\mathrm{cm}^2/\mathrm{V}^2$ and $\alpha=-70\,\hbar\,\mathrm{MHz}\,\mathrm{cm}^2/\mathrm{V}^2$, respectively. Dashed lines indicate second-order polynomial fits to the numerical results.}
    \label{fig:anhrmonic_effect}
\end{figure}

The equations of motion derived from the complete Lagrangian form a system of three coupled nonlinear differential equations that, in general, cannot be solved analytically. The equations were solved numerically using a standard ordinary differential equation solver implemented in Python. Since we are interested in the COM mode, the initial position vector was chosen to be proportional to the corresponding eigenvector $\vec{v}_1$ amplitudes $A(1,1,\frac{a}{\sqrt{r_m}})$ while all initial velocities were set to zero. $A$ is an arbitrary amplitude of the mode (normalized by the length scale) and $a$ is the relative atom participation as defined for $\vec{v}_1$ before.\\
The resulting trajectories were Fourier transformed, and the COM oscillation frequency was extracted from the dominant spectral peak. Since the motion is mostly harmonic, the frequency peaks are mostly separated and
correspond to the frequencies of the normal modes. Repeating the procedure for different oscillation amplitudes yields the dependence of the COM frequency on the motional amplitude.\\
The resulting frequency dependence on the oscillation amplitude is shown in Fig.~\ref{fig:anhrmonic_effect} for both positive and negative atomic polarizabilities. We immediately see that the frequency shift tends to increase with the amplitude. A quadratic polynomial was fitted to the numerical data and used to estimate the temperature-dependent frequency correction. The resulting scaling for positive polarizability is given by,
\begin{equation}
\sqrt{\lambda_1}\left(A,\alpha=\alpha_{\min}\right)
\approx
0.99 + 0.0012A - 0.153A^2,
\end{equation}

and for negative polarizability,
\begin{equation}
\sqrt{\lambda_1}\left(A,\alpha=\alpha_{\min}\right)
\approx
1.011 + 0.0038A - 0.239A^2.
\end{equation}

\newpage

\bibliographystyle{unsrt}
\bibliography{references}

\end{document}